**Neglecting Model Structural Uncertainty Underestimates Upper Tails of Flood Hazard**

Tony E. Wong[1,†,*], Alexandra Klufas[2], Vivek Srikrishnan[3], Klaus Keller[1,4,5]

[1]Earth and Environmental Systems Institute, Pennsylvania State University, University Park, PA 16802, USA.
[2]Department of Mathematics, Wellesley College, Wellesley, MA 02481, USA.
[3]Department of Energy and Mineral Engineering, Pennsylvania State University, University Park, PA 16802, USA.
[4]Department of Geosciences, Pennsylvania State University, University Park, PA 16802, USA.
[5]Department of Engineering and Public Policy, Carnegie Mellon University, Pittsburgh, PA 15289, USA.

† Now at: Department of Computer Science, University of Colorado, Boulder, CO 80309, USA.

* Corresponding author: Tony Wong, anthony.e.wong@colorado.edu

**Abstract**
Coastal flooding drives considerable risks to many communities, but projections of future flood risks are deeply uncertain. The paucity of observations of extreme events often motivates the use of statistical approaches to model the distribution of extreme storm surge events. One key deep uncertainty that is often overlooked is model structural uncertainty. There is currently no strong consensus among experts regarding which class of statistical model to use as a "best practice". Robust management of coastal flooding risks requires coastal managers to consider the distinct possibility of non-stationarity in storm surges. This increases the complexity of the potential models to use, which tends to increase the data required to constrain the model. Here, we use a Bayesian model averaging approach to analyze the balance between (i) model complexity sufficient to capture decision-relevant risks and (ii) data availability to constrain complex model structures. We characterize deep model structural uncertainty through a set of calibration experiments. Specifically, we calibrate a set of models ranging in complexity using long-term tide gauge observations from the Netherlands and the United States. We find that in both considered cases, roughly half of the model weight is associated with the non-stationary models. Our approach provides a formal framework to integrate information across model structures, in light of the potentially sizable modeling uncertainties. By combining information from multiple models, our inference sharpens for the projected storm surge 100-year return levels, and estimated return levels increase by several centimeters. We assess the impacts of data availability through a set of experiments with temporal subsets and model comparison metrics. Our analysis suggests that about 70 years of data are required to stabilize estimates of the 100-year return level, for the locations and methods considered here.

**1 Introduction**

Storm surges drive substantial risks to coastal communities (Nicholls and Cazenave 2010), but there remains deep structural uncertainty regarding how best to model this threat. Previous work has broken important new ground by considering process-based modeling (Fischbach *et al* 2017, Orton *et al* 2016, Johnson *et al* 2013) as well as statistical modeling approaches (Buchanan *et al* 2015, Grinsted *et al* 2013, Tebaldi *et al* 2012, Menéndez and Woodworth 2010). Recently, we have seen the advent of semi-empirical models for sea-level rise and their application to coastal risk management (Kopp *et al* 2017, Nauels *et al* 2017, Wong *et al* 2017a, 2017b, Mengel *et al* 2016). The total flood hazard depends on predictions of both sea-level rise and storm surge properties. In this case, it can be attractive to have flexible and efficient models to estimate storm surge hazards, with a formal statistical accounting of uncertainty and linked to accessible climate variables. This motivates our study's focus on the statistical modeling of storm surges.

Previous studies have provided important new insights by examining the potentially sizable impacts of non-stationarity in the treatment of storm frequency, distribution and intensity (e.g., Ceres *et al* 2017, Lee *et al* 2017, Cid *et al* 2016, Grinsted *et al* 2013, Haigh *et al* 2010b, Menéndez and Woodworth 2010). For example, Grinsted *et al* (2013) use a generalized extreme value (GEV) distribution to model extreme sea levels, and incorporate non-stationarity in the model parameters by allowing them to covary with global mean surface temperature. Other studies consider a hybrid statistical model wherein the frequency of extreme sea level events is governed by a Poisson process (PP) and the magnitude of these events follows a Generalized Pareto distribution (GPD) (Wahl *et al* 2017, Hunter *et al* 2017, Buchanan *et al* 2017, Cid *et al* 2016, Bulteau *et al* 2015, Marcos *et al* 2015, Arns *et al* 2013, Tebaldi *et al* 2012). Non-stationarity may be incorporated into the PP/GPD statistical model by covarying the PP/GPD parameters with climatic conditions (Marcos *et al* 2015, Haigh *et al* 2010b). Here, we follow and expand on the work of Haigh *et al* (2010b) and examine how non-stationarity – covarying with changing North Atlantic oscillation (NAO) index – affects projections of future storm surge return levels using a PP/GPD model.

Extreme events are, by definition, rare. It is hence important to use the relatively sparse data well. The GEV approach requires to bin observations into time blocks, processed in a manner so as to remove the interdependence of the observations, and take block maxima. Often, this is done using annual blocks (e.g., Wong and Keller 2017, Karamouz *et al* 2017), yielding a potentially limited amount of data with which to fit an extreme value statistical model (Coles 2001). Another option is to process data to achieve independence, then use shorter time lengths of blocks (Grinsted *et al* 2013), but the choice of processing procedure is nontrivial and the fidelity with which non-stationary behavior may be detected is uncertain (e.g., Ceres *et al* 2017, Lee *et al* 2017). The PP/GPD modeling approach is an attractive option because all events above a specified threshold are considered in fitting the model, leading to a richer set of data (e.g., Knighton *et al* 2017, Arns *et al* 2013). While we do not employ these methods, it is important to note that other approaches exist to analyze extreme sea levels; for example, those based on the joint probability method (McMillan *et al* 2011, Haigh *et al* 2010a, Tawn and Vassie 1989, Pugh and Vassie 1978). Previous work has examined how data availability affects model prediction (Dangendorf *et al* 2016), but this question remains largely open for longer tide gauge records (>90 years).

A related open question is how to select a statistical model of extreme storm surges. Relative to stationary models, the increased complexity of non-stationary models can lead to wider predictive intervals, and perhaps the dismissal of the more complex model – along with arguably decision-relevant tail behavior. Traditional approaches often favor parsimonious use of the limited data (e.g., Karamouz *et al* 2017, Lee *et al* 2017, Buchanan *et al* 2015, Tebaldi *et al* 2012). Bayesian model averaging (BMA), however, offers an avenue to combine a range of candidate model structures by allowing the data to inform the degree to which each model is to be trusted (Hoeting *et al* 1999). Models are a proxy for data not yet observed, and our BMA approach presents an opportunity to formally integrate multiple information streams (Moftakhari *et al* 2017).

Here, we combine the non-stationarity covarying with NAO index with a PP/GPD modeling approach to address the interrelated questions of how data length affects model choice, and how model choice impacts estimates of storm surge hazards. We employ the PP/GPD model because we are motivated by the need to examine how best to utilize the inherently limited data regarding extreme sea levels. We use two relatively long and complete tide gauge records to demonstrate that for both sites and all data lengths, non-stationary models receive considerable weight in a Bayesian model averaging sense. The major contributions of this study are: (i) to present a formal statistical framework to combine information across models and account for structural uncertainties through use of Bayesian model averaging, and (ii) to assess how the length of data record affects our model choices, and thus impacts estimates of future flood hazard.

## 2 Methods

### 2.1 Storm surge statistical modeling

We employ a peaks-over-thresholds (POT) approach, with a PP/GPD statistical model, to estimate the distribution of extreme storm surge events. We find similar conclusions in an experiment assessing the implications of our results using a block maxima approach in the region considered by Grinsted *et al* (2013) (see supplementary material). The POT approach makes use of only observational data that exceed a specified threshold to fit the PP/GPD model parameters. We follow previous work (e.g., Wahl *et al* 2017, Arns *et al* 2013) and process the data by: (i) using a constant threshold *μ(t)* equal to the 99th percentile of the daily maximum water levels, (ii) detrending by subtracting a moving window one-year average from the raw hourly data (or three-hourly for Delfzijl) to account for sea-level rise but retain sub-decadal variability, the effects of astronomical tides, and interannual variability, as well as the effects of storm surges, and (iii) using a declustering routine to isolate extreme events at least 72 hours apart. In coastal risk management applications, these methods would be used together with a set of local mean sea-level rise projections that would likely have an annual time step. Thus, it is important to retain these non-mean sea level signals. In a set of supplemental experiments, we also examine a declustering time-scale of 24 hours and POT thresholds of the 95th and 99.7th percentiles. The interested reader is referred to Arns *et al* (2013) for a careful review of key structural uncertainties.

The probability density function (pdf, *f*) and cumulative distribution function (cdf, *F*) for the potentially non-stationary form of the GPD used here are given by:

$$f(x(t); \mu(t), \sigma(t), \xi(t)) = \frac{1}{\sigma(t)}\left(1 + \xi(t)\frac{x(t)-\mu(t)}{\sigma(t)}\right)^{-(1/\xi(t)+1)} \quad (1)$$

$$F(x(t); \mu(t), \sigma(t), \xi(t)) = 1 - \left(1 + \xi(t)\frac{x(t)-\mu(t)}{\sigma(t)}\right)^{-1/\xi(t)}, \quad (2)$$

where *x(t)* is the processed daily maximum tide gauge level (meters), *σ(t)* is the scale parameter (meters) and *ξ(t)* is the shape parameter (unitless), all as functions of time *t* (days). *σ* governs the width of the distribution and *ξ* governs the heaviness of the distribution's tail. A Poisson process governs the probability *g* of observing *n(t)* exceedances of threshold *μ(t)* during time interval Δ*t* (days):

$$g(n(t); \lambda(t)) = \frac{(\lambda(t)\Delta t)^{n(t)}}{n(t)!} \exp(-\lambda(t)\Delta t), \quad (3)$$

where *λ(t)* is the Poisson rate parameter (exceedances day$^{-1}$).

We incorporate potential non-stationarity into the PP/GPD model following the approach of Grinsted *et al* (2013), by allowing the model parameters to covary with winter (DJF) average NAO index:

$$\begin{cases} \lambda(t) = \lambda_0 + \lambda_1 NAO(t) \\ \sigma(t) = \exp[\sigma_0 + \sigma_1 NAO(t)] \\ \xi(t) = \xi_0 + \xi_1 NAO(t). \end{cases} \quad (4)$$

$\lambda_0$, $\lambda_1$, $\sigma_0$, $\sigma_1$, $\xi_0$, and $\xi_1$ are uncertain model parameters, determined by fitting to the processed tide gauge record (detailed below). We assume the parameters are stationary within each year. The processing of tide gauge data into a surge index in Grinsted *et al* (2013) serves to (1) achieve independence among observations, and (2) increase the effective amount of data by pooling across sites. Regarding (1), we process our tide gauge data to achieve independence (see above). Regarding (2), we are investigating how data availability affects our ability to constrain storm surge statistical models, and what the impacts are on model projections relevant to managing local coastal risks. We use direct tide gauge data instead of a surge index because we are currently unaware of any method to map surge index back to a localized projection.

Finally, the joint likelihood function for the model parameters ***θ*** = ($\lambda_0$, $\lambda_1$, $\sigma_0$, $\sigma_1$, $\xi_0$, $\xi_1$)$^T$, given the time series of daily maxima threshold exceedances, *x*, is:

$$L(x|\boldsymbol{\theta}) = \prod_{i=1}^{N} \left[ g(n(y_i); \lambda(y_i)) \prod_{j=1}^{n(y_i)} f(x_j(y_i); \mu(y_i), \sigma(y_i), \xi(y_i)) \right], \tag{5}$$

where $i = 1, 2, \ldots, N$ indexes the years of tide gauge data and $j = 1, 2, \ldots, n(y_i)$ indexes the exceedances $x_j(y_i)$ in year $y_i$. The product indexed by $j$ in equation (5) is replaced by 1 for all $i$ such that $n(y_i) = 0$.

We consider four candidate models within the class of PP/GPD models, ranging from a stationary model (denoted by "ST", in which $\lambda_1 = \sigma_1 = \xi_1 = 0$) to fully non-stationary ("NS3", in which all six parameters are considered). These models are summarized in table 1. We project future storm surge return levels to 2065. We focus on the 100-year return level, which is motivated by its common use in coastal risk management (e.g., Coastal Protection and Restoration Authority of Louisiana 2017), but results for other return periods are presented in the supplementary material.

| Model Structure | Non-stationary Parameters | Model Parameters to Calibrate |
|---|---|---|
| ST | None | $\lambda_0, \sigma_0, \xi_0$ |
| NS1 | $\lambda$ | $\lambda_0, \lambda_1, \sigma_0, \xi_0$ |
| NS2 | $\lambda, \sigma$ | $\lambda_0, \lambda_1, \sigma_0, \sigma_1, \xi_0$ |
| NS3 | $\lambda, \sigma, \xi$ | $\lambda_0, \lambda_1, \sigma_0, \sigma_1, \xi_0, \xi_1$ |

**Table 1.** Candidate model structures and their parameters.

## 2.2 Model calibration

### 2.2.1 Data

We fit the candidate models' parameters (table 1) using the tide gauge data record from two sites: Delfzijl, the Netherlands (Rijkswaterstaat 2017), and Sewells Point (Norfolk), Virginia, United States (NOAA 2017). We selected these sites because the lengths of the records (137 and 89 years, respectively) enable our set of experiments regarding the impacts of data length on surge level estimation, they are geographically well-separated and these tide gauge records are relatively complete (each site has three or fewer gaps longer than one month). We use time series of detrended daily block maxima for the POT approach (e.g., Arns *et al* 2013).

We use historical monthly NAO index data from Jones *et al* (1997). We use the sea level pressure projection of the MPI-ECHAM5 simulation under SRES scenario A1B as part of the ENSEMBLES project (www.ensembles-eu.org; Roeckner *et al* 2003). We calculate the winter mean (DJF) NAO index following Stephenson *et al* (2006) to use as input to the nonstationary models. We caution that these results do not account for model structural nor parametric uncertainty regarding future NAO index. An assessment of the impacts of these uncertainties on projected surge levels is another important avenue for future study.

We evaluate the impacts of data length on PP/GPD parameter estimates through a set of experiments. In these experiments, we employ only the 30, 50, 70, 90, 110 and 137 most recent years of data from the Delfzijl tide gauge site, and the 30, 50, 70 and 89 most recent years from Norfolk.

### 2.2.2 Bayesian calibration framework

We calibrate each of the four candidate models (table 1) using each of the two processed tide gauge records ($x(t)$) and winter NAO index series ($NAO(t)$). We employ a robust adaptive Markov chain Monte Carlo approach (Vihola 2012). The essence of this calibration approach is to update the prior probability distribution of the model parameters ($p(\boldsymbol{\theta})$) by quantifying the goodness-of-fit between the observational data and the Poisson process/generalized Pareto models given by candidate sets of model parameters. This goodness-of-fit is quantified by the likelihood function (equation 5). Bayes' theorem combines the prior knowledge regarding the model parameters with the information gained from the observational data (i.e., the likelihood function) into the posterior distribution of the model parameters, given the data ($p(\boldsymbol{\theta}|x)$):

$$p(\boldsymbol{\theta}|x) \propto L(x|\boldsymbol{\theta})\, p(\boldsymbol{\theta}). \tag{6}$$

We represent prior knowledge regarding the parameters ($p(\boldsymbol{\theta})$) as follows. First, we obtain maximum likelihood parameter estimates (MLEs) for 28 tide gauge sites with at least 90 years of data available, as well as the two records on which this study focuses. These sites were selected using the University of Hawaii Sea Level Center's online database, and a spreadsheet utility we developed (and provide with the model codes in the repository accompanying this study) (Caldwell *et al* 2015). Details regarding these sites are provided in the supplementary material accompanying this article. Second, we fit either a normal or gamma distribution to the set of 30 MLEs for each parameter, depending on whether the parameter has infinite (normal: $\lambda_1, \sigma_1, \xi_0, \xi_1$) or half-infinite (gamma: $\lambda_0, \sigma_0$) support. The resulting prior distributions, MLEs, and an experiment using uniform prior distributions are shown in the supplementary material.

We initialize the Markov chains at the MLE parameters for each site and for each candidate model. We produce 500,000 iterations for 10 parallel Markov chains and remove the first 50,000 iterations for burn-in. Gelman and Rubin diagnostics are used to assess convergence and burn-in length (Gelman and Rubin 1992). For each site, for each of the four candidate models, and for each of the data length experiments, we draw an ensemble of 10,000 parameter sets for analysis from the remaining 4,500,000 Markov chain samples. We calibrate in this manner for each of the length of data experiments (see sect. 2.2.1).

We also conduct a preliminary experiment by binning the Delfzijl data into 11 overlapping 30-year blocks, spanning the 137-year range. We calibrate the stationary model (ST) to the data in each of the 11 blocks, and calculate the estimated 100-year return level for each block's ensemble. We examine changes in the quantiles of these 11 distributions to assess the potential need for a non-stationary approach.

**2.2.3 Bayesian model averaging**

Bayesian model averaging (BMA) (Hoeting *et al* 1999) is a method by which the storm surge return level estimates implied by the posterior parameters (obtained as in section 2.2.2) for each candidate model (table 1) may be combined and weighted by the model marginal likelihood, given the data, $p(M_k|x)$. Let $RL(y_i|x,M_k)$ denote the return level in year $y_i$ assuming model structure $M_k \in \{ST, NS1, NS2, NS3\}$ and given the observational data $x$. Then the BMA-weighted return level in year $y_i$, integrating the estimates from all four candidate models, is

$$RL(y_i|x) = \sum_{k=1}^{4} RL(y_i|M_k)\, p(M_k|x). \tag{7}$$

The BMA weights, $p(M_k|x)$, are given by

$$p(M_k|x) = \frac{p(x|M_k)\,p(M_k)}{\sum_{l=1}^{4} p(x|M_l)\,p(M_l)}, \qquad (8)$$

where the denominator marginalizes the probability of the data, *p(x)*, over the four model structures considered. We make the assumption that all model structures are equally likely *a priori* (i.e., $p(M_k)$ = $p(M_l)$, $\forall M_k, M_l \in$ {*ST, NS1, NS2, NS3*}. The probabilities *p(x|M_k)* are determined by integration over the posterior distributions of the model parameters:

$$p(x|M_k) = \int_{\boldsymbol{\theta}} p(x|\boldsymbol{\theta}, M_k)\,p(\boldsymbol{\theta})\,d\boldsymbol{\theta}, \qquad (9)$$

where the integral is over the relevant parameters for model $M_k$. The probabilities *p(x|θ,M_k)* are the likelihood function (equation 5) with conditional dependence on the model structure made explicit. These and the prior probabilities (*p(θ)*) are sampled as described in section 2.2.2.

From equation (9), *p(x|M_k)* is the normalizing constant (or *marginal likelihood*) for the probability density function associated with model $M_k$. We use bridge sampling (Meng and Wing 1996) to estimate the marginal likelihoods of the models under consideration, using a normal approximation to the joint posterior as the importance density.

**2.2.4 Model comparison metrics**

We employ several metrics for model comparison. They are motivated by the balance between model goodness-of-fit, model complexity, and the availability of data. The first metric is the Akaike information criterion (AIC) (Akaike 1974):

$$AIC = -2\log(L_{max}) + 2N_p, \qquad (10)$$

where $L_{max}$ is the maximum value of the likelihood function (equation 5) within the posterior model ensemble and $N_p$ is the number of model parameters.

The second metric is the Bayesian information criterion (BIC) (Schwarz 1978):

$$BIC = -2\log(L_{max}) + N_p \log(N_{obs}), \qquad (11)$$

where $N_{obs}$ is the number of observational data used to fit the model. Thus, for $N_{obs} > e^2$, BIC penalizes overparameterization more harshly than AIC.

The third metric is the deviance information criterion (DIC) (Spiegelhalter *et al* 2002). For a given model structure, define the *deviance* for a given set of model parameters as *D(θ)* = -2 log(*L(x|θ)*). Denote by $\bar{D}$ the expected value of *D(θ)* over *θ*, and let $\bar{\boldsymbol{\theta}}$ refer to the expected value of *θ*. The *effective number of parameters* is calculated as $p_D = \bar{D} - D(\bar{\boldsymbol{\theta}})$. DIC is then:

$$DIC = p_D + \bar{D}. \qquad (12)$$

The final metric we employ for model comparison is the BMA weights themselves (equation 8). Note that AIC and BIC are calculated based on the performance of the maximum likelihood ensemble member, whereas DIC and BMA weight are based on the entire ensemble.

In addition to the four ensembles corresponding to each of the candidate models (table 1), we construct a BMA-weighted ensemble of estimated return levels as follows. We draw 10,000 sets of parameters from each of the four candidate models. The number of samples was selected to match the number of samples used for each individual model. For each of these 10,000 concomitant sets of BMA parameters, we calculate the return period according to equation (7).

## 3 Results

### 3.1 Hindcast test

The Delfzijl site displays evidence for non-stationary behavior in the 100-year return level (figure 1). We determine the distributions shown in figure (1) by binning the data into overlapping 30-year blocks and fitting the stationary (ST) model using the Bayesian approach outlined in sect. 2.2.2. The estimated median 100-year return level ranges from 412 to 490 cm across the 11 blocks, and widths of the 5-95% credible interval range from 146 to 285 cm. This motivates the need for a non-stationary approach.

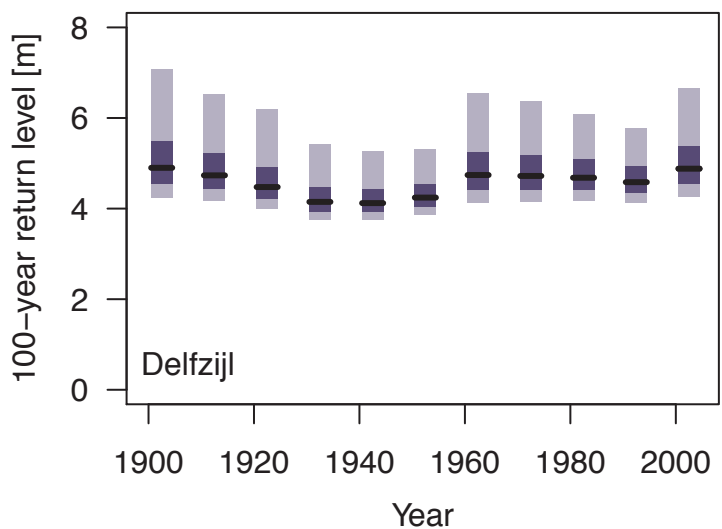

**Figure 1.** Distributions of the 100-year return levels at Delfzijl, the Netherlands, using the stationary (ST) model and 30-year blocks of data, centered at the locations of the vertical bars. The black horizontal lines denote the ensemble median; the dark and light bars denote the 25-75% and 5-95% quantile ranges, respectively.

We find that the more complex models (NS1, NS2 and NS3) generally result in somewhat lower traditional model performance metrics (table 2). However, we note that differences of *O(1)* in AIC or BIC may not be sufficient evidence to dismiss the more complex models (Kass and Raftery 1995). For both sites, the BMA weights associated with the non-stationary models NS1 and NS2 are roughly between 20 and 30%, indicating the value of multi-model approaches over single-model or stationary modeling approaches.

| Tide Gauge | Model Structure | AIC | BIC | DIC | BMA weight |
|---|---|---|---|---|---|
| Delfzijl, the Netherlands | ST | 5545.62 | 5557.26 | 13855.07 | 0.42 |
| | NS1 | 5546.07 | 5561.59 | 13852.99 | 0.33 |
| | NS2 | 5547.70 | 5567.10 | 13853.27 | 0.22 |
| | NS3 | 5548.21 | 5571.50 | 13851.72 | 0.04 |
| Norfolk, Virginia, USA | ST | 2883.20 | 2893.21 | 7198.87 | 0.51 |
| | NS1 | 2884.39 | 2897.74 | 7198.76 | 0.24 |
| | NS2 | 2886.27 | 2902.96 | 7199.84 | 0.20 |
| | NS3 | 2886.73 | 2906.75 | 7198.11 | 0.05 |

**Table 2.** Model selection criteria for the four candidate models. Lower is better for AIC, BIC and DIC; higher is better for BMA weight. Shaded cells denote the model choice indicated by each metric.

### 3.2 Estimates of current and future surge levels

The resulting predictive distributions for 2016 and projected 2065 surge levels demonstrate the impacts of integrating across model structures (figure 2; see supplementary material for these results in tabular form). Interestingly, the NS3 model displays a *reduction* in 100-year return level for both sites by 2065, but also receives the lowest BMA weight (about 5%). The fact that the ST, NS1 and NS2 models' projections are in relative agreement and match the data well (see table 2) lends confidence to their results. This agreement, characterized by quite similar posterior pdfs, leads to a tighter credible range in the BMA projection (figure 2). While the sharpened inference in the BMA pdf in this case may seem counterintuitive, this follows from the fact that the BMA return levels are *averages* of the return levels from the four candidate models. Averaging is a smoothing operation, so extreme behavior is dampened (see also supplementary material for a note describing this phenomenon). Indeed, a key strength of our BMA approach is to formally quantify the degree of belief in each model structure, informed by the quality of model match to data.

We find that a stationary PP/GPD approach underestimates projected 100-year surge levels in 2065 by 3 and 4 cm for Delfzijl and Norfolk, respectively, relative to the BMA approach (ensemble medians, figure 2c, d; see also tables S2 and S3). While 3 cm may not seem like a substantial increase in hazard, it is ultimately up to the decision-maker to assess the relevant hazards for themselves, and our BMA approach incorporates model specification uncertainty into the projections presented. In any case, these results serve as a proof of concept of the use of Bayesian model averaging in a statistical treatment of extreme sea levels, and characterize the model structural uncertainty.

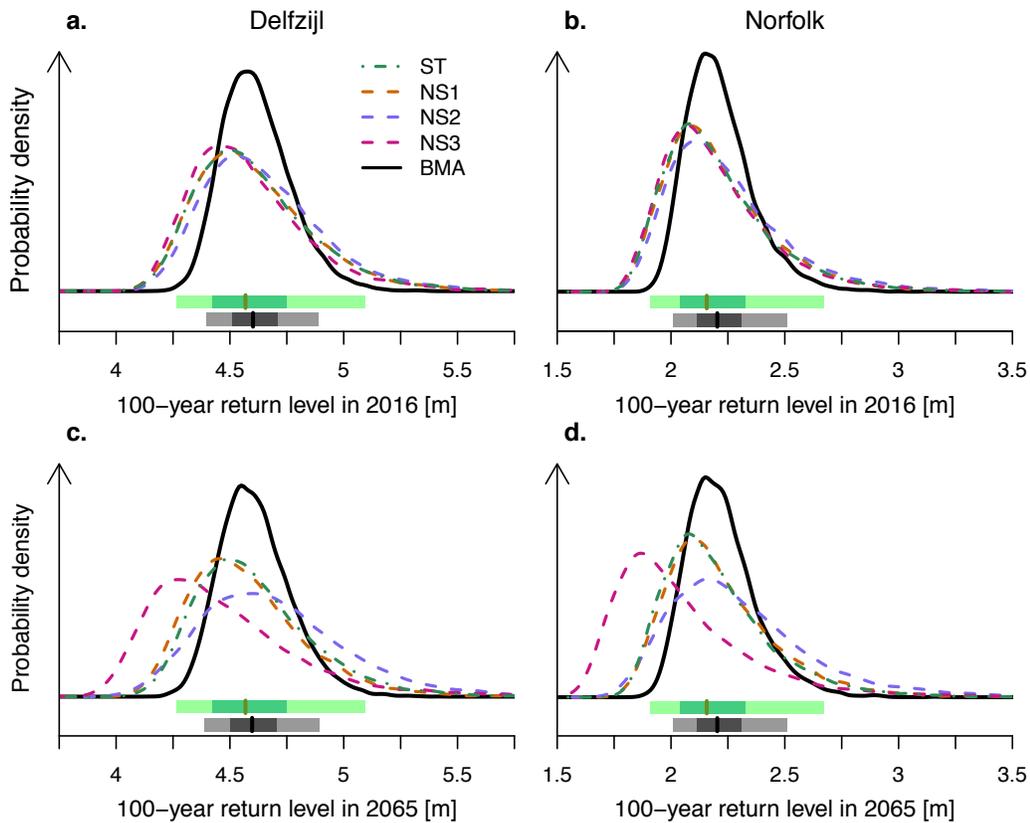

**Figure 2.** Posterior probability density function (pdfs) of the 100-year return levels using the BMA-weighted ensemble and four individual model structures for 2016 conditions (top row) and projected 2065 surge levels (bottom row), at Delfzijl (left column) and Norfolk (right column). Below the pdfs are boxplots for the stationary model (ST, green horizontal bars) and the BMA-weighted ensemble (gray/black horizontal bars). The bold vertical lines denote the ensemble medians; the dark and light bars denote the 25-75% and 5-95% quantile ranges, respectively.

### 3.3 Reliability of estimated surge levels

We assess the impacts of data length on the distributions of PP/GPD parameters for the four candidate models (figure 3). With a relatively short record (30-50 years of data), 5-95% credible intervals for the 100-year surge level are much wider than when 70 or more years of data are available. While it is beyond the scope of this study, future work might consider developing a formal convergence metric using (for example) Kolmogorov and Smirnov statistics (Smirnov 1948, Kolmogorov 1933).

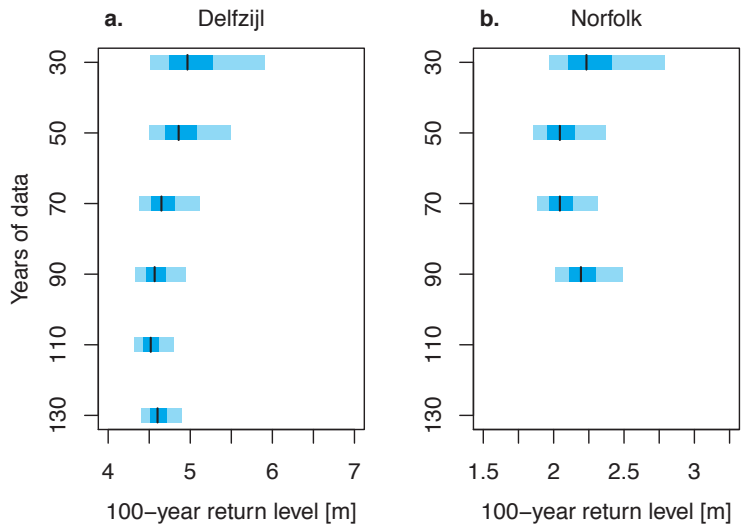

**Figure 3.** Distributions of return levels from the BMA-weighted ensemble in year 2016 for varying time lengths of data employed for (a) Delfzijl and (b) Norfolk. The black vertical lines denote the ensemble medians; the extent of the darker boxes gives the 25-75% credible range; the extent of the lighter boxes gives the 5-95% credible range.

The BMA weights change for each site as more data become available, but once 70 years of data are available, the ordering of the models' BMA weights remains stable (figure 4). Across all of the data length experiments, the stationary model has the largest BMA weight for both sites, at about 40-50%. As more data become available at Delfzijl, the stationary model receives less than 50% weight and models NS1 and NS2 receives roughly 30 and 20% weight, respectively. We find similar results for Norfolk. It is consistent and clear across sites and data lengths, however, that the non-stationary models receive about half of the model weight. This result is also robust to changes in the selected POT threshold (see supplementary material). This illustrates the potential limitations of single-model approaches.

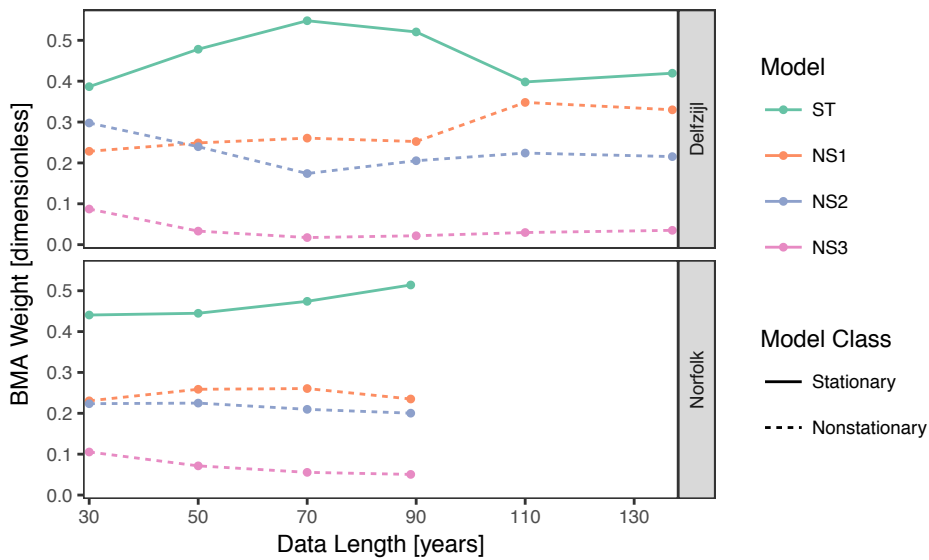

**Figure 4.** Bayesian model averaging weights (equation 8) for the four candidate models, using varying lengths of data from the tide gauge stations at Delfzijl (137 years of data total) and Norfolk (89 years). Higher values imply better model-data match.

## 4 Discussion

Our analysis (i) showcases a new framework to integrate decision-relevant information (i.e., non-stationarity) into storm surge projections and (ii) uses this framework to demonstrate practical implications of neglecting key modeling uncertainties. Our analysis, of course, is subject to several caveats. For example, our BMA approach weights each model according to its posterior probability (under a uniform prior over the model space), thereby implicitly using a quadratic loss function with respect to the choice of model (Robert 2007). The quadratic loss function may not be the most appropriate loss for all applications using storm surge distributions. A fruitful future study might assess the impacts of alternative loss functions tailored to specific decision problems. Additionally, other applications may require sampling approaches other than the bridge sampling employed here (e.g., Yao *et al* 2017).

We caution that our analysis focuses on NAO index as a covariate for the storm surge statistical model parameters, but there may well be other useful predictors for modulating surge. It is a, perhaps, counterintuitive result that two sites on opposite sides of the Atlantic display similar behavior in storm surge return levels response to changing NAO index (c.f., Figure 2) when one might expect to see opposite effects for the two sites. We hypothesize that this is due to the fact that in our simple single covariate model, any non-stationarity must be attributed to NAO index. Future work might consider incorporating other potential predictors, to test for additional drivers of storm surge non-stationarity.

We focus on the 100-year return level, but provide results for other return periods in the supplementary material. Higher return levels likely require more data for the same constraint, and fewer data for lower return levels. Furthermore, we find tighter constraint on the 100-year return level in the BMA ensemble, as a result of convergent projections from the four candidate models. This may not always be the case, and implementing our BMA approach with more diverse sets of candidate models is an important avenue for future work. Combining information across model structures using BMA can be of use in decision-making by more efficiently integrating the available information, and tighter constraint on projected flood hazard can help to avoid potential over-/under-protection regrets. Finally, many previous statistical treatments of storm surge hazard have made a single-model assumption (e.g., Grinsted *et al* 2013). Our results suggest that this may yield an overestimate of the range in projected flood hazard, so it is important to formally assess the impacts of those assumptions.

## 5 Conclusions

We present a framework for incorporating model structural uncertainty into estimates of coastal surge level probability, using two long tide gauge records to evaluate the impacts of data availability on our results. Our analysis indicates that previous work using a stationary Poisson process/generalized Pareto distribution modeling approach may underestimate the upper tails of flood hazards, and overestimate the uncertainty range. Discarding models on the basis of performance metrics (table 2) or by assuming a single model structure neglects model structural uncertainty that may be captured through our BMA approach. Our results highlight the impacts of neglecting key modeling uncertainties on estimates of storm surge return levels, and are of practical use to provide a more complete picture of decision-relevant information for the management of coastal flood risks.

**Acknowledgments**

We gratefully acknowledge Benjamin Lee, Alexander Bakker, David Johnson, Jordan Fischbach, Debra Knopman, Neil Berg, Nathan Urban, Ryan Sriver, Nancy Tuana, Robert Lempert, Murali Haran, Robert

Ceres, Robert Nicholas, Chris Forest and Jesse Nusbaumer for valuable inputs. This work was co-supported by the National Science Foundation through the Network for Sustainable Climate Risk Management (SCRiM) under NSF cooperative agreement GEO-1240507; the National Oceanic and Atmospheric Administration (NOAA) through the Mid-Atlantic Regional Integrated Sciences and Assessments (MARISA) program under NOAA grant NA16OAR4310179; and the Penn State Center for Climate Risk Management. The ENSEMBLES data used in this work was funded by the EU FP6 Integrated Project ENSEMBLES (Contract number 505539) whose support is gratefully acknowledged.

We are not aware of any real or perceived conflicts of interest for any authors. Any conclusions or recommendations expressed in this material are those of the authors and do not necessarily reflect the views of the funding agencies. Any errors and opinions are, of course, those of the authors. All results, model codes, analysis codes, data and model outputs used for analysis are freely available from https://github.com/tonyewong/MESS and are distributed under the GNU general public license. The datasets, software tools and other resources on this website are provided as-is without warranty of any kind, express or implied. In no event shall the authors or copyright holders be liable for any claim, damages or other liability in connection with the use of these resources.
**References**

Akaike H 1974 A New Look at the Statistical Model Identification *IEEE Trans. Automat. Contr.* **19** 716–23

Arns A, Wahl T, Haigh I D, Jensen J and Pattiaratchi C 2013 Estimating extreme water level probabilities: A comparison of the direct methods and recommendations for best practise *Coast. Eng.* **81** 51–66

Buchanan M K, Kopp R E, Oppenheimer M and Tebaldi C 2015 Allowances for evolving coastal flood risk under uncertain local sea-level rise *Clim. Change* **137** 347–62

Buchanan M K, Oppenheimer M and Kopp R E 2017 Amplification of flood frequencies with local sea level rise and emerging flood regimes *Environ. Res. Lett.* **12** 64009

Bulteau T, Idier D, Lambert J and Garcin M 2015 How historical information can improve estimation and prediction of extreme coastal water levels: Application to the Xynthia event at la Rochelle (France) *Nat. Hazards Earth Syst. Sci.* **15** 1135–47

Caldwell P C, Merrfield M A and Thompson P R 2015 Sea level measured by tide gauges from global oceans - the Joint Archive for Sea Level holdings (NCEI Accession 0019568), Version 5.5 *NOAA Natl. Centers Environ. Information, Dataset*

Ceres R, Forest C E and Keller K 2017 Understanding the detectability of potential changes to the 100-year peak storm surge *Clim. Change* **145** 221–35

Cid A, Menéndez M, Castanedo S, Abascal A J, Méndez F J and Medina R 2016 Long-term changes in the frequency, intensity and duration of extreme storm surge events in southern Europe *Clim. Dyn.* **46** 1503–16

Coastal Protection and Restoration Authority of Louisiana 2017 *Louisiana's Comprehensive Master Plan for a Sustainable Coast* (Baton Rouge, LA) Online: http://coastal.la.gov/our-plan/2017-coastal-master-plan/

Coles S G 2001 *An introduction to Statistical Modeling of Extreme Values*

Dangendorf S, Arns A, Pinto J G, Ludwig P and Jensen J 2016 The exceptional influence of storm "Xaver" on design water levels in the German Bight *Environ. Res. Lett.* **11**

Fischbach J R, Johnson D R and Molina-Perez E 2017 *Reducing Coastal Flood Risk with a Lake Pontchartrain Barrier* (Santa Monica, CA, USA) Online: https://www.rand.org/pubs/research_reports/RR1988.html

Gelman A and Rubin D B 1992 Inference from Iterative Simulation Using Multiple Sequences *Stat. Sci.*
13

**Supplemental Figures and Tables**
Accompanying "Neglecting Model Structural Uncertainty Underestimates Upper Tails of Flood Hazard"
Authors: Wong et al.

| University of Hawaii record number | Location | Country | Latitude (° N) | Longitude (° E) | Record length (years) |
|---|---|---|---|---|---|
| 57 | Honolulu | United States | 21.3 | -157.86 | 110.0 |
| 571 | Ketchikan | United States | 55.33 | -131.63 | 96.2 |
| 540 | Prince Rupert | Canada | 54.32 | -130.32 | 106.0 |
| 542 | Tofino | Canada | 49.15 | -125.92 | 105.2 |
| 543 | Victoria | Canada | 48.42 | -123.37 | 105.9 |
| 551 | San Francisco (Fort Point) | United States | 37.8 | -122.47 | 117.4 |
| 567 | Los Angeles | United States | 33.72 | -118.27 | 91.1 |
| 554 | La Jolla | United States | 32.87 | -117.25 | 90.2 |
| 569 | San Diego | United States | 32.71 | -117.17 | 108.9 |
| 775 | Galveston (Pier 21) | United States | 29.33 | -94.74 | 111.0 |
| 762 | Pensacola | United States | 30.40 | -87.21 | 91.7 |
| 242 | Key West | United States | 24.55 | -81.81 | 101.9 |
| 240 | Fernandina Beach | United States | 30.68 | -81.47 | 117.6 |
| 261 | Charleston | United States | 32.78 | -79.93 | 93.2 |
| 266 | Cristobal | Panama | 9.37 | -79.88 | 107.7 |
| 302 | Balboa | Panama | 8.97 | -79.57 | 107.5 |
| 264 | Atlantic City | United States | 39.35 | -74.42 | 103.4 |
| 745 | New York | United States | 40.7 | -74.02 | 94.6 |
| 741 | Boston | United States | 42.35 | -71.05 | 93.7 |
| 252 | Portland | United States | 43.66 | -70.25 | 104.8 |
| 275 | Halifax | Canada | 44.68 | -63.61 | 118.2 |
| 294 | Newlyn | United Kingdom | 50.1 | -5.54 | 95.7 |
| 822 | Brest | France | 48.38 | -4.5 | 169.0 |
| 824 | Marseille | France | 43.28 | 5.35 | 165.2 |
| 825 | Cuxhaven | Germany | 53.87 | 8.72 | 97.0 |
| 837 | Gedser | Denmark | 54.57 | 11.93 | 121.3 |
| 838 | Hornbaek | Denmark | 56.1 | 12.47 | 122.0 |
| 826 | Stockholm | Sweden | 59.32 | 18.08 | 126.0 |
| - | **Sewells Point (Norfolk)** | **United States** | **36.95** | **76.33** | **89** |
| - | **Delfzijl** | **Netherlands** | **53.33** | **6.93** | **137** |

**Table S1**. Tide gauge stations where maximum likelihood parameter estimates were used to fit the prior distributions. Bold-faced text denotes the two sites on which the analysis presented in the main text focuses.

| Model structure | 2016 [m] | | | | | | | 2065 [m] | | | | | | |
|---|---|---|---|---|---|---|---|---|---|---|---|---|---|---|
| | Min. | 5% | 25% | 50% | 75% | 95% | Max. | Min. | 5% | 25% | 50% | 75% | 95% | Max. |
| ST | 4.05 | 4.27 | 4.42 | 4.57 | 4.75 | 5.10 | 6.24 | 4.05 | 4.27 | 4.42 | 4.57 | 4.75 | 5.10 | 6.24 |
| NS1 | 4.02 | 4.26 | 4.42 | 4.57 | 4.75 | 5.09 | 6.28 | 4.00 | 4.23 | 4.39 | 4.53 | 4.71 | 5.04 | 6.23 |
| NS2 | 3.96 | 4.29 | 4.46 | 4.61 | 4.79 | 5.14 | 7.34 | 3.94 | 4.25 | 4.47 | 4.66 | 4.88 | 5.29 | 7.94 |
| NS3 | 4.04 | 4.24 | 4.40 | 4.54 | 4.71 | 5.05 | 6.29 | 3.86 | 4.07 | 4.23 | 4.40 | 4.64 | 5.12 | 8.45 |
| BMA | 4.17 | 4.40 | 4.51 | 4.60 | 4.71 | 4.89 | 5.37 | 4.15 | 4.39 | 4.50 | 4.60 | 4.71 | 4.89 | 5.42 |

**Table S2.** Quantiles of the estimated 100-year surge level (meters) for Delfzijl in 2016 and in 2065.

| Model structure | 2016 [m] | | | | | | | 2065 [m] | | | | | | |
|---|---|---|---|---|---|---|---|---|---|---|---|---|---|---|
| | Min. | 5% | 25% | 50% | 75% | 95% | Max. | Min. | 5% | 25% | 50% | 75% | 95% | Max. |
| ST | 1.73 | 1.91 | 2.04 | 2.16 | 2.32 | 2.67 | 4.16 | 1.73 | 1.91 | 2.04 | 2.16 | 2.32 | 2.67 | 4.16 |
| NS1 | 1.74 | 1.90 | 2.04 | 2.16 | 2.33 | 2.67 | 4.15 | 1.74 | 1.91 | 2.05 | 2.17 | 2.35 | 2.70 | 4.11 |
| NS2 | 1.76 | 1.92 | 2.06 | 2.19 | 2.37 | 2.75 | 4.35 | 1.64 | 1.90 | 2.08 | 2.25 | 2.47 | 2.93 | 4.92 |
| NS3 | 1.73 | 1.89 | 2.03 | 2.15 | 2.31 | 2.66 | 4.86 | 1.55 | 1.70 | 1.84 | 1.98 | 2.19 | 2.66 | 4.70 |
| BMA | 1.88 | 2.01 | 2.11 | 2.19 | 2.30 | 2.49 | 3.19 | 1.84 | 2.01 | 2.11 | 2.20 | 2.31 | 2.51 | 3.19 |

**Table S3.** Quantiles of the estimated 100-year surge level (meters) for Norfolk in 2016 and in 2065.

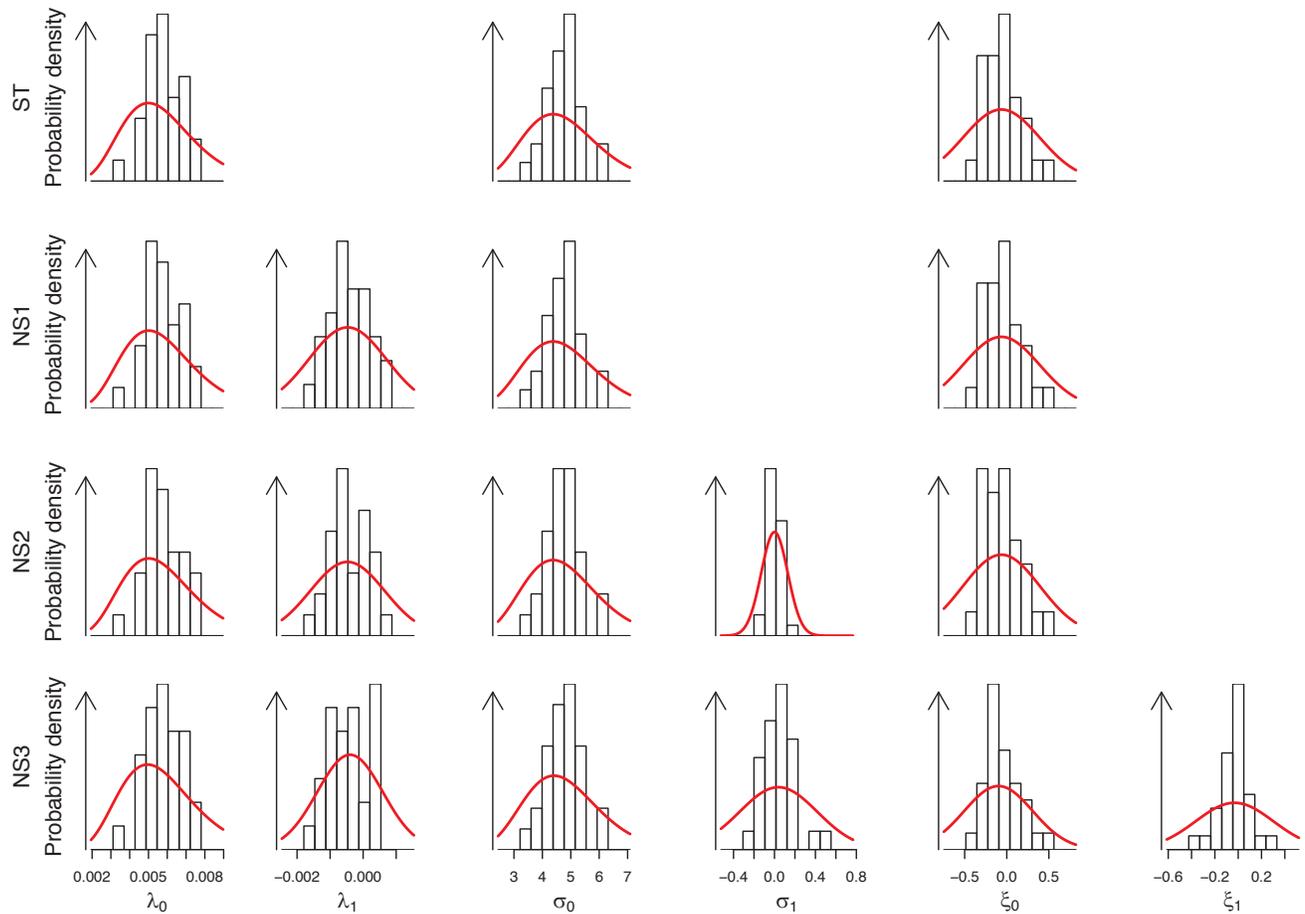

**Figure S1.** Histograms of the maximum likelihood parameter estimates for the network of 30 tide gauge sites (see table S1), with the fitted normal or gamma prior distributions superimposed.

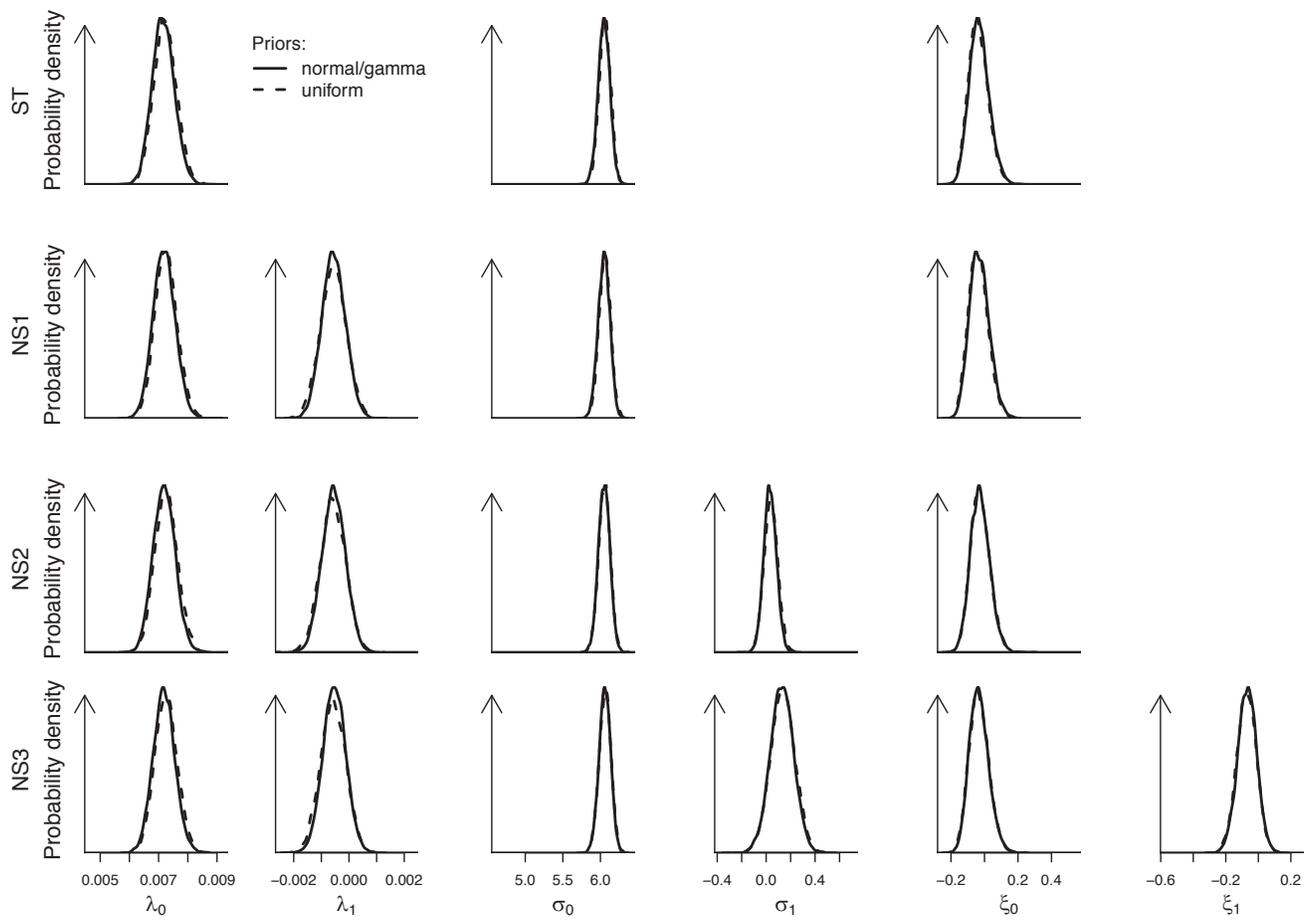

**Figure S2.** Posterior distributions of model parameters for Delfzijl, the Netherlands, using as prior distributions either wide uniform distributions (dashed lines) or the fitted normal/gamma prior distributions from figure S1 (solid lines).

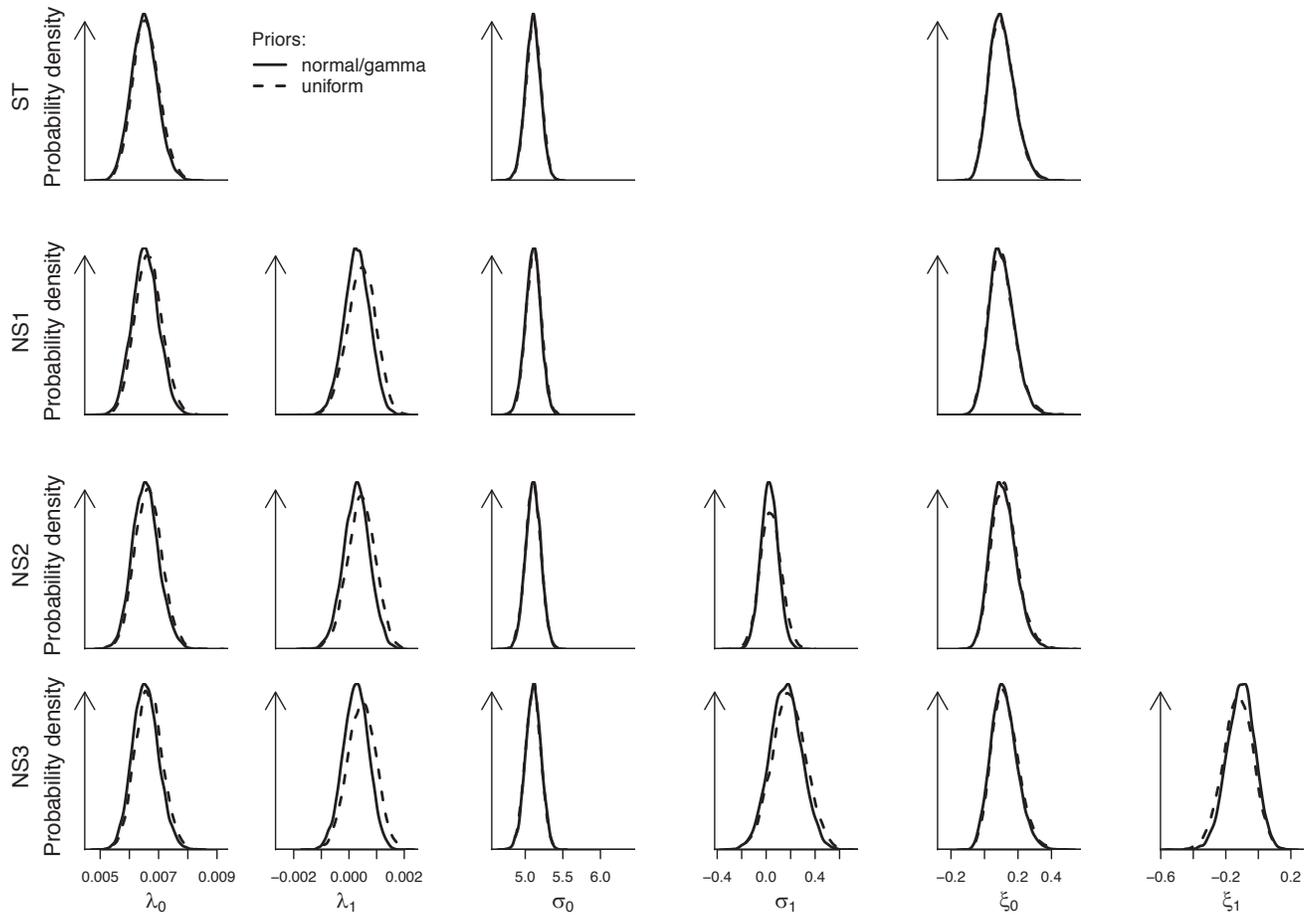

**Figure S3.** Posterior distributions of model parameters for Sewells Point (Norfolk), United States, using as prior distributions either wide uniform distributions (dashed lines) or the fitted normal/gamma prior distributions from figure S1 (solid lines).

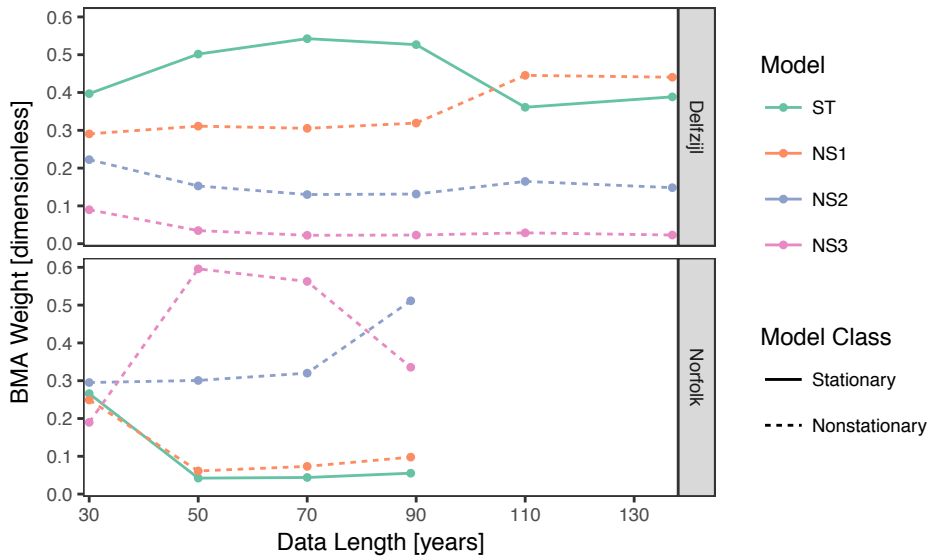

**Figure S4.** Bayesian model averaging weights (main text, equation 8) for the four candidate models, using varying lengths of data from the tide gauge stations at Delfzijl (137 years of data total) and Norfolk (89 years) and the 95th quantile as the peaks-over-thresholds cutoff (main text uses the 99th quantile). Higher weight values imply better model-data match.

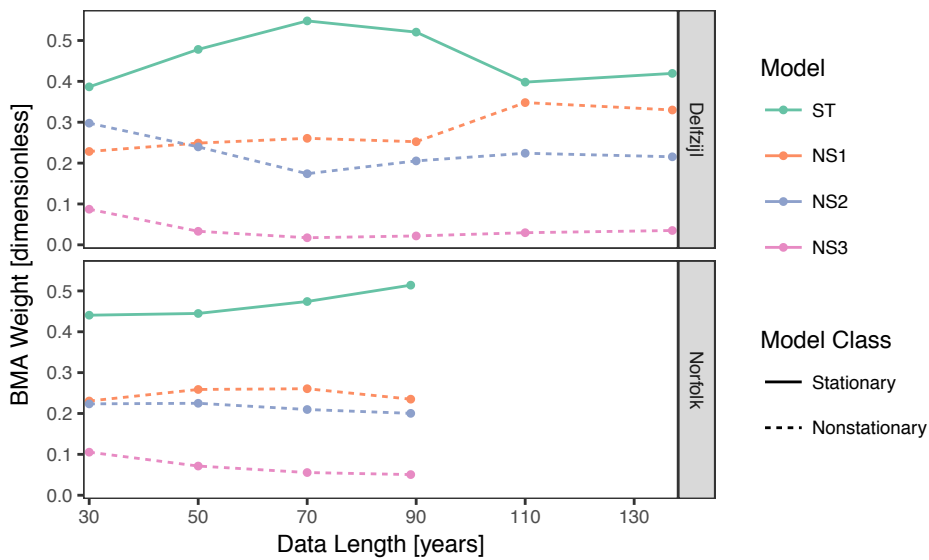

**Figure S5.** Bayesian model averaging weights (main text, equation 8) for the four candidate models, using varying lengths of data from the tide gauge stations at Delfzijl (137 years of data total) and Norfolk (89 years) and the 99.7th quantile as the peaks-over-thresholds cutoff (main text uses the 99th quantile). Higher weight values imply better model-data match.

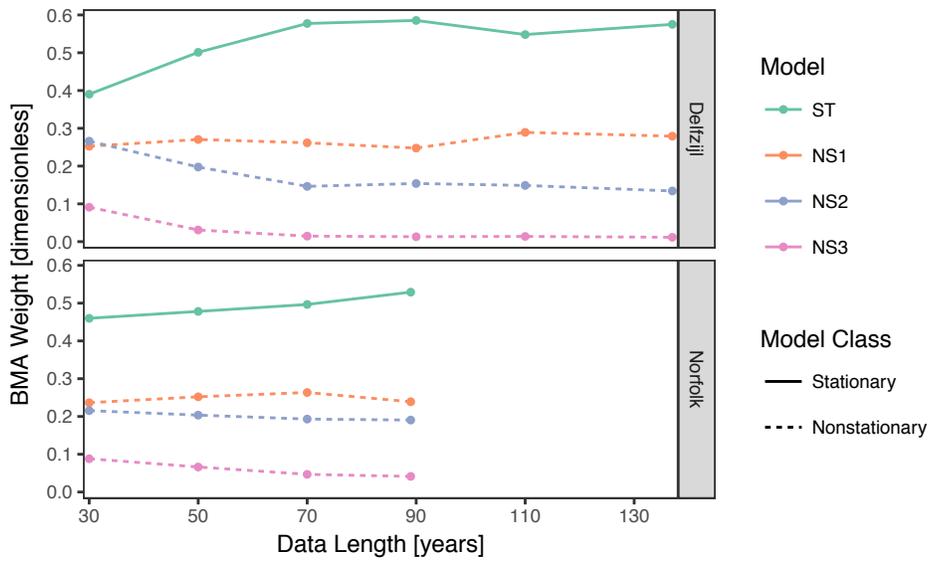

**Figure S6.** Bayesian model averaging weights (main text, equation 8) for the four candidate models, using varying lengths of data from the tide gauge stations at Delfzijl (137 years of data total) and Norfolk (89 years) and a 1-day declustering time-scale (main text uses a 3-day declustering time-scale). Higher weight values imply better model-data match.

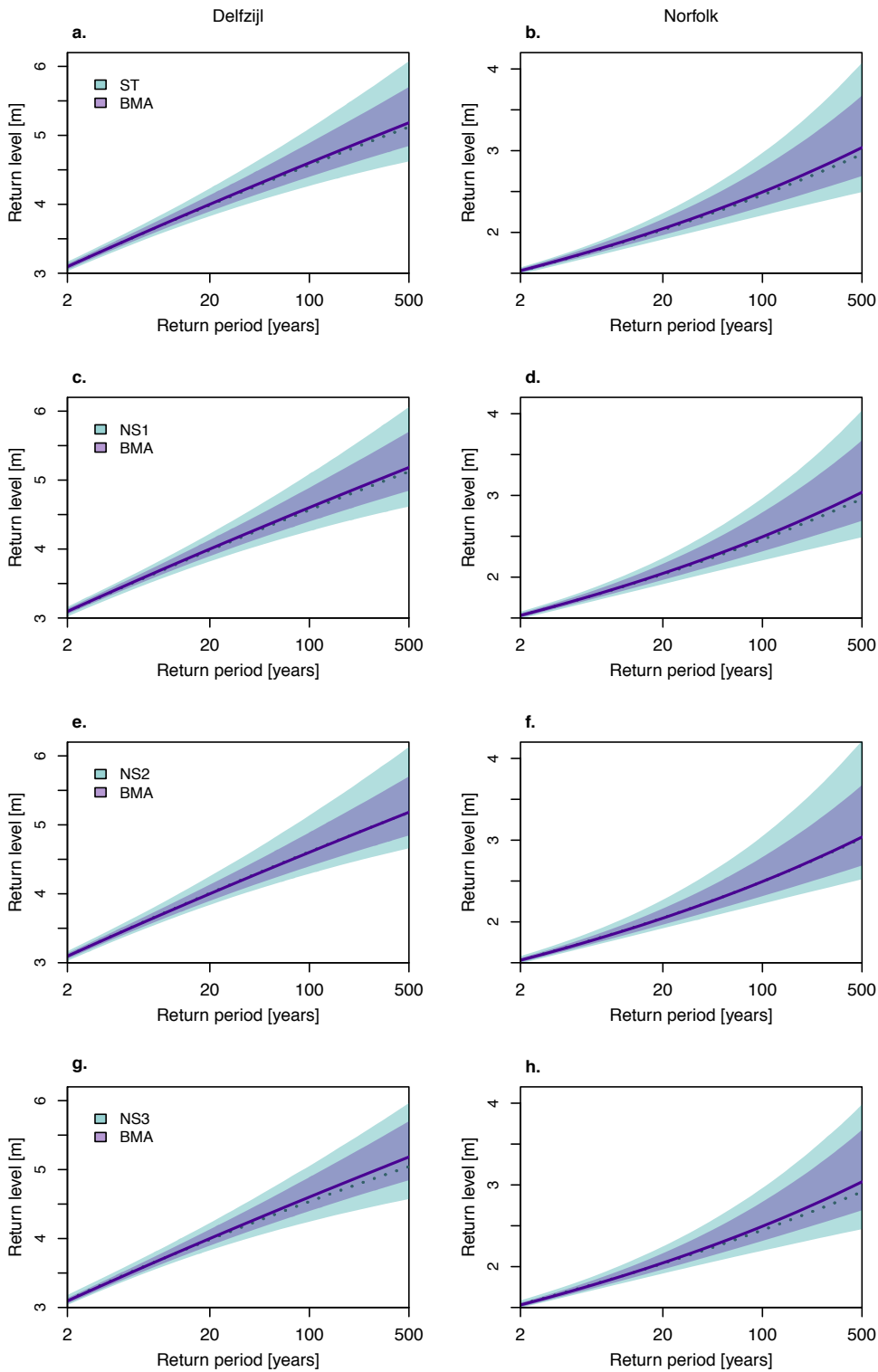

**Figure S7.** Storm surge return periods (years) and associated return levels (meters) in 2016 for Delfzijl (left column) and Norfolk (right column), for the Bayesian model average-weighted ensemble and stationary model (top row), NS1 model (second row), NS2 model (third row) and NS3 model (bottom row).

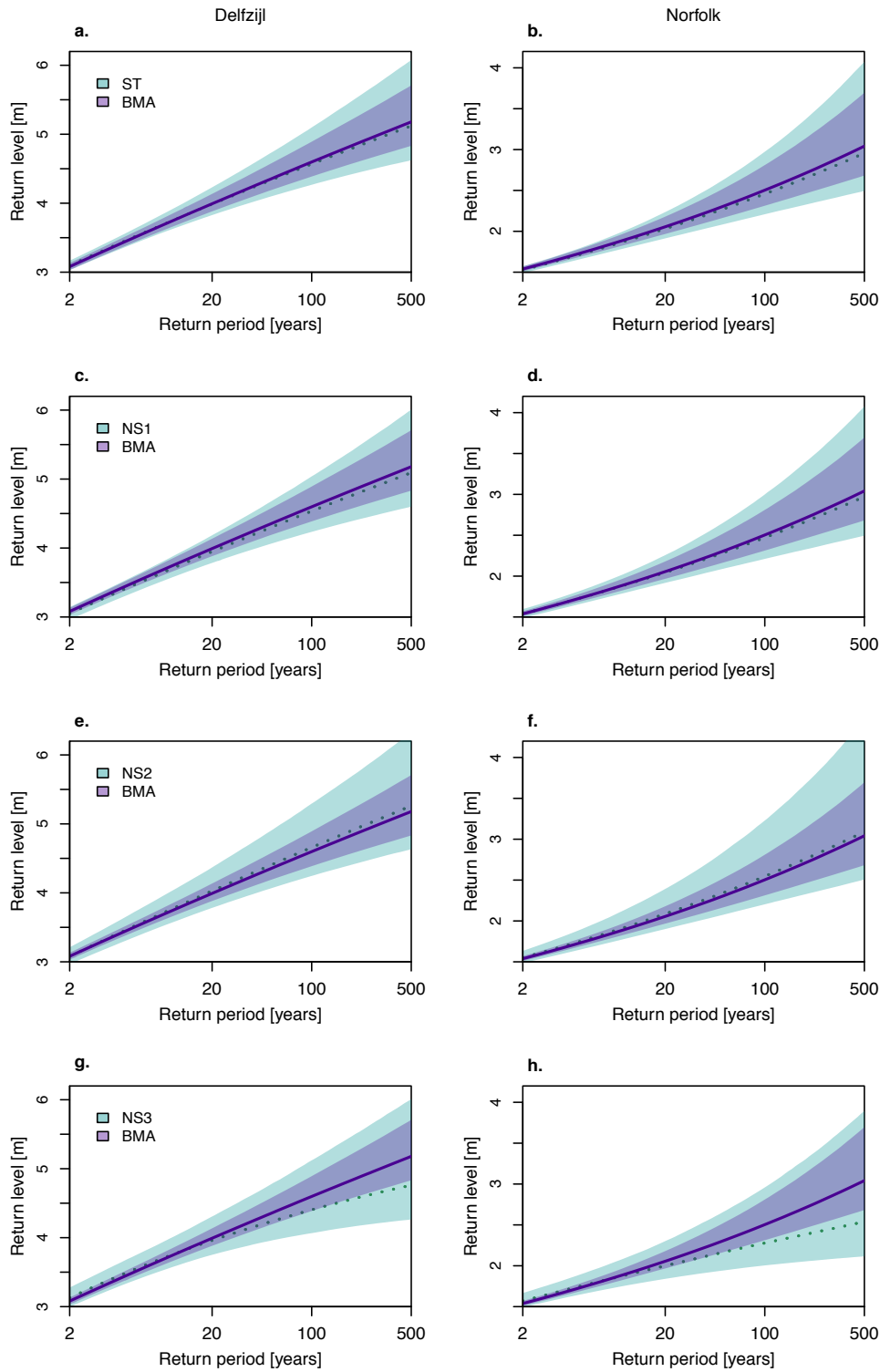

**Figure S8.** Storm surge return periods (years) and associated return levels (meters) in 2065 for Delfzijl (left column) and Norfolk (right column), for the Bayesian model average-weighted ensemble and stationary model (top row), NS1 model (second row), NS2 model (third row) and NS3 model (bottom row).

**Supplemental Text 1**
Accompanying "Neglecting Model Structural Uncertainty Underestimates Upper Tails of Flood Hazard"
Authors: Wong et al.

This supplementary text describes a supplemental experiment conducted to assess the impacts of model structural uncertainties and sensitivity to data length of return level estimates based on a generalized extreme value distribution modeling approach.

**Methods**

**Observational data and case study**
We use data from four long record tide gauge stations along the East Coast of the United States (table 1). We obtained tide data from the University of Hawaii Sea Level Center data portal (Caldwell *et al* 2015). We detrend the observations by subtracting each year's annual mean from the record to account for sea-level rise, then take the time series of detrended annual block maxima for analysis. In our preprocessing, we remove any years of observational data missing more than 10% of the year's total data.

**Table 1.** Tide stations used in study.

| Tide Station | Latitude, Longitude | Observational Record (Years) | Years Used in Study | Total Number of Missing Days | Largest Gap (Days) |
|---|---|---|---|---|---|
| Atlantic City, New Jersey | 39.35° N, -74.42° W | 103 | 91 | 1037 | 556 |
| Boston, Massachusetts | 42.35° N, -71.05° W | 93 | 91 | 53 | 9 |
| New London, Connecticut | 41.37° N, -72.10° W | 76 | 66 | 201 | 30 |
| Portland, Maine | 43.66° N, -70.25° W | 104 | 90 | 113 | 14 |

**Statistical modeling of extreme storm surges**
A generalized extreme value (GEV) distribution is the limiting distribution for a series of block maxima. Three parameters - location ($\mu$), scale ($\sigma$), and shape ($\xi$) - govern the character of the GEV probability density function:

$$f(x \mid \mu, \sigma, \xi) = \frac{1}{\sigma} z(x)^{\xi+1} e^{-z(x)} \tag{1}$$

$$z(x) = \begin{cases} (1 + \xi(\frac{x-\mu}{\sigma}))^{\frac{-1}{\xi}} & if\ \xi \neq 0 \\ e^{\frac{-(x-\mu)}{\sigma}} & if\ \xi = 0 \end{cases} \tag{2}$$

We follow recent work (Grinsted *et al* 2013, Lee *et al* 2017) and the main text by allowing each GEV model parameter to covary with global mean surface temperature:

$$\mu(t) = \mu_0 + \mu_1\, T(t) \tag{3}$$

$$\sigma(t) = e^{\sigma_0 + \sigma_1 T(t)} \tag{4}$$

$$\xi(t) = \xi_0 + \xi_1\, T(t). \tag{5}$$

A traditional stationary GEV model would have $\mu_1 = \sigma_1 = \xi_1 = 0$ in equations 3-5, and thus have three free parameters. A fully non-stationary GEV would follow equations 3-5 in their entirety and have six free parameters. We follow the guidance of Lee *et al* (2017) and the main text and consider a range of candidate GEV model structures ranging from fully stationary to fully non-stationary (table 2).

**Table 2.** Candidate models.

| Model Name | Non-Stationary Parameters | Estimated Parameters |
|---|---|---|
| ST | None | $\mu_0, \sigma_0, \xi_0$ |
| NS1 | $\mu$ | $\mu_0, \mu_1, \sigma_0, \xi_0$ |
| NS2 | $\mu, \sigma$ | $\mu_0, \mu_1, \sigma_0, \sigma_1, \xi_0$ |
| NS3 | $\mu, \sigma, \xi$ | $\mu_0, \mu_1, \sigma_0, \sigma_1, \xi_0, \xi_1$ |

**Model calibration**
We calculate maximum likelihood parameter estimates using a differential evolution algorithm (Storn and Price 1997). The joint likelihood function for a non-stationary GEV distribution is

$$f(x \mid \mu, \sigma, \xi) = \prod_{i=1}^{N} \frac{1}{\sigma(t_i)}\, z(x(t_i))^{\xi(t_i)+1} e^{-z(x(t_i))}, \tag{6}$$

where $z(x)$ is given by equation 2, $i=1, 2, 3, \ldots, N$ indexes the number of data points (i.e., $N$ is the number of annual block maxima in the time series), and $x(t)$ is the time series of annual block maxima. We use the log-likelihood over the traditional likelihood function (equation 6) for numerical stability. We fit maximum likelihood estimates for each model structure (table 2) to each tide data set (table 1).

**Sensitivity to length of observational record**
In a series of data length sensitivity experiments, we artificially limit each data set to the most recent 10, 20, 30, … years of tide gauge data. We calculate the sensitivity of each parameter (we use $\theta$ to denote a generic parameter) of each candidate model by comparing the parameter estimate with the assimilated data set artificially limited to $t$ years in length ($\theta^t$), to the parameter estimate with all available data ($\theta$):

$$\Delta \theta^t = \frac{|\theta^t - \theta|}{\theta}. \tag{7}$$

The closer the value is to zero, the more closely the parameter estimate reflects the parameter estimate of the entire data set. Thus, values of $\Delta \theta^t$ close to zero imply that $t$ years of data are representative of the entire data set, as far as the calculated parameter estimates are concerned.

Similarly, we quantify the impacts of data length on the estimated 20-year flood return levels:

$$\Delta RL^t = \frac{|RL^t - RL|}{RL}, \tag{8}$$

where *RL* is the return level (meters) of the 20-year flood as estimated by maximum likelihood using all of the data and $RL^t$ is the estimate calculated using a data set limited to the most recent *t* years of data. Positive values of $\Delta RL^t$ indicate an underestimation of the height of the 20-year flood and negative values indicate an overestimation of the height of the 20-year flood, as a result of experimentally limiting the supply of tide gauge data. We select to examine the 20-year flood here so there is sizable probability that at least one "true" 20-year flood event has been observed in a tide gauge record of 60 years or longer (less than 5% probability that in 60 years no 20-year flood events are observed).

**Results and discussion**

We find that the maximum likelihood parameter estimates require at least about 60 years of data before the estimates based on artificially-limited data are representative of the estimates calculated using the complete data set (figure 1). In line with expectations, we note that the non-stationary parameters, $\mu_1$, $\sigma_1$ and $\xi_1$, are the most difficult to constrain (note the magnitude of the vertical scales in figure 1). We find that $\xi$ requires the most data to estimate reliably. This is also not surprising in light of the fact that $\xi$ governs the weight of the tails of the GEV distribution. This finding is in line with the results of (Lee *et al* 2017). Similarly, we find that estimates of the 20-year return level require about 60 years of data in order for the artificially data-limited flood risk estimates to reliably represent those from the complete data set (figure 2). The Bayesian calibration method applied in the main text is more sophisticated than the maximum likelihood approach employed in this supplemental experiment, and the main text focused on the 100-year flood. Thus, it is expected that more data will be required to stabilize those 100-year flood return level estimates. In light of this, these results (60 years minimum data) are roughly in line with the conclusions presented in the main text for stabilization of return level estimates calculated using a Bayesian calibration and a Poisson process/generalized Pareto distribution approach (70 years minimum data).

There are, of course, some caveats surrounding this simple analysis. For example, we reduce the number of data points available to fit the extreme value model by processing our data to yearly block maxima. If we were to use smaller time blocks, such as monthly block maxima, in our preprocessing, we may obtain different results using generalized extreme value distributions regarding projections and parameter stabilization (Lee *et al* 2017, Grinsted *et al* 2013). Tide gauge location may also play a role in our results and conclusions. We focus only on tide gauge stations along the East Coast of the U.S., thus these results may not be representative of the rest of the world. Future work should consider a broader range of tide gauge locations. We have also focused this analysis on the 20-year flood. Future work should also consider the data requirements to constrain estimates for less frequent floods, such as the 50 or 100-year flood.

In spite of these caveats, we find that the parameter and return level estimates begin to stabilize with at least 60 years of available data (figures 1 and 2). These results can provide practical guidance for future use of extreme value statistical models to project storm surge risk, and sheds light on potential biases and limitations of this common approach to managing coastal flood risk.

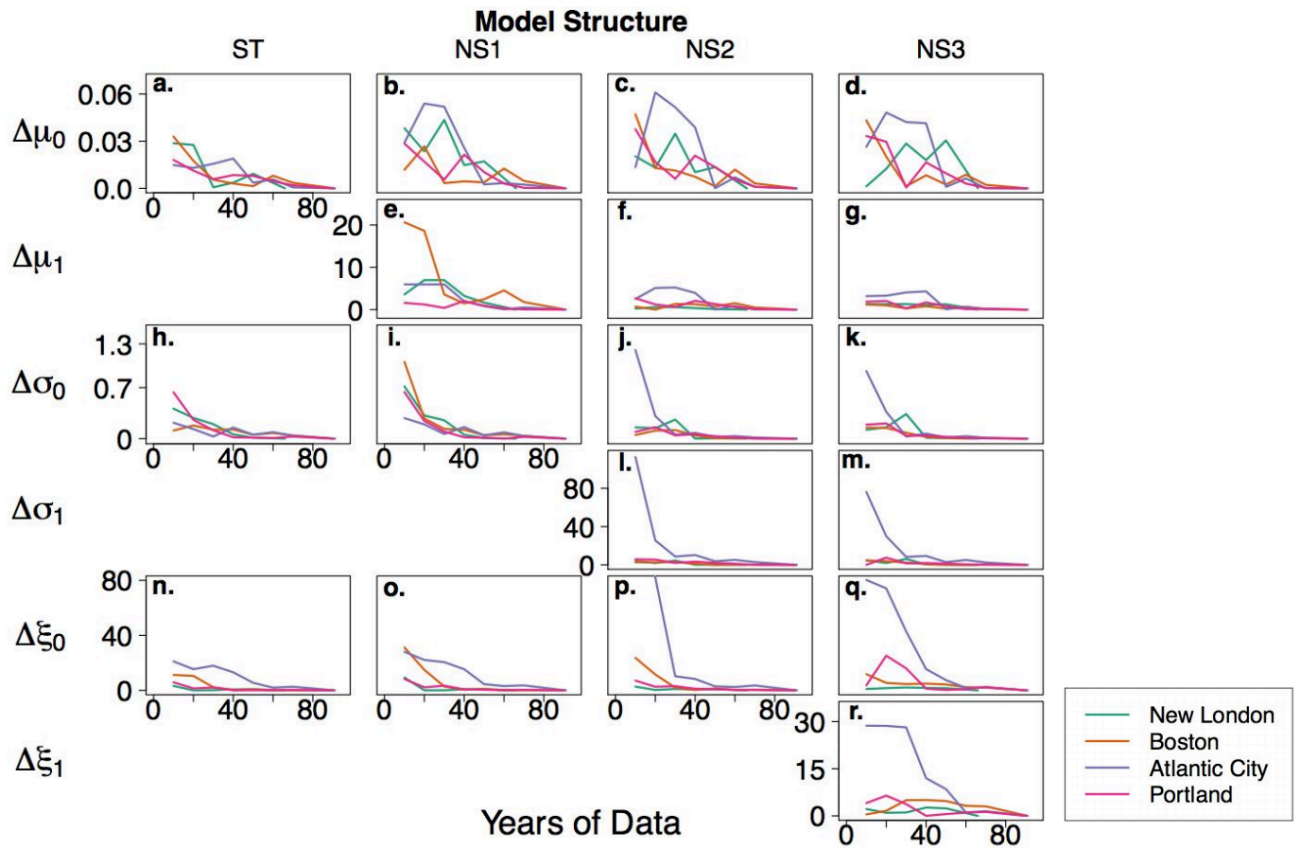

**Figure 1.** Stabilization of parameter estimates with increase in data length for each of the model parameters (rows), four candidate model structures (columns) and four tide gauge sites (time series).

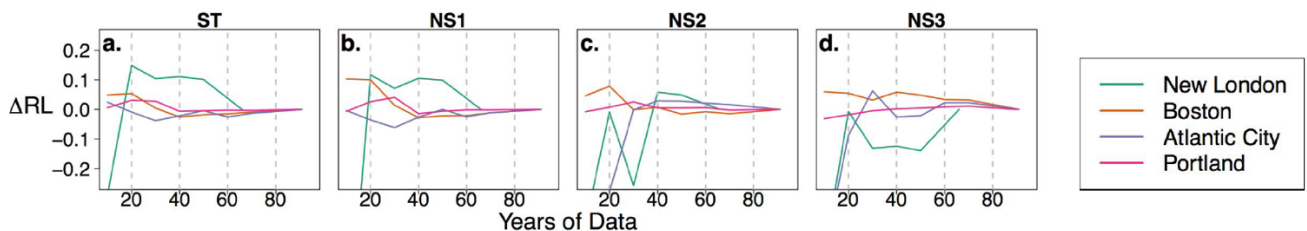

**Figure 2.** Comparison of the 20-year flood return level with increase in data length for the four candidate model structures and four tide gauge sites.

**Supplemental Text 2**
Accompanying "Neglecting Model Structural Uncertainty Underestimates Upper Tails of Flood Hazard"
Authors: Wong et al.

This supplementary text provides a brief interpretation of the narrowing of distributions under the Bayesian model averaging (BMA) in the main text, which may seem counterintuitive. We demonstrate in a simplified case that the BMA-weighted ensemble should have a distribution which has lower variance – i.e., it is narrower – than the distributions of the candidate models (ST, NS1, NS2 and NS3).

We make the simplifying but illustrative approximation that the distributions for the 100-year return level under each candidate model (Table 1 and Figure 2, main text) are independent and follow normal distributions. This makes the algebra tractable for this demonstration. For clarity, we only consider two candidate models (call them $X$ and $Y$). So, let the distributions for $X$ and $Y$ be as follows,

$$X \sim N(\mu_x, \sigma_x^2)$$
$$Y \sim N(\mu_y, \sigma_y^2),$$

where $\mu_x$ and $\mu_y$ are the means of their distributions, and $\sigma_x^2$ and $\sigma_y^2$ are their variances. Then, the distribution of $X+Y$ is the convolution of the distributions for $X$ and $Y$ as above, which is

$$X + Y \sim N(\mu_x + \mu_y, \sigma_x^2 + \sigma_y^2).$$

Suppose the BMA weights of the models are $w_x$ and $w_y$, such that $w_x+w_y=1$ and $0 \leq w_x, w_y \leq 1$ (since the BMA weights are constitute a probability distribution). Then, the BMA-weighted ensemble is generated as

$$Z = w_x X + w_y Y.$$

Furthermore, using standard results from probability theory, the distribution of the BMA-weighted ensemble has mean and variance given by:

$$\mu_z = w_x \mu_x + w_y \mu_y$$
$$\sigma_z^2 = w_x^2 \sigma_x^2 + w_y^2 \sigma_y^2.$$

Let $\sigma_m^2 = \max(\sigma_x^2, \sigma_y^2)$. Then:

$$\sigma_z^2 = w_x^2 \sigma_x^2 + w_y^2 \sigma_y^2 \leq w_x^2 \sigma_m^2 + w_y^2 \sigma_m^2 \leq \sigma_m^2 (w_x^2 + w_y^2) \leq \sigma_m^2,$$

where the final inequality holds because of the constraints $w_x+w_y=1$ and $0 \leq w_x, w_y \leq 1$. Thus, the distribution of the estimated flood levels from the BMA-weighted ensemble must necessarily have lower variance than the candidate models' return level estimates. Equality holds only in the trivial case where one model carries all of the BMA weight.